\documentclass[preprint,12pt]{elsarticle}

\usepackage{amsmath}
\usepackage{bm}
\usepackage{amssymb}
\usepackage{graphicx}
\usepackage{caption}
\usepackage{floatrow}
\usepackage{subcaption}
\usepackage{graphicx,xcolor} 
\usepackage[hidelinks]{hyperref}
\usepackage{array}
\usepackage{float}

\usepackage[normalem]{ulem}

\newcommand{\<}{\langle}
\renewcommand{\>}{\rangle}

\journal{Computational Materials Science}

\begin{document}
\begin{frontmatter}



\title{Clarifying the Ti–V Phase Diagram Using First-Principles Calculations and Bayesian Learning}

\author[label1]{Timofei Miryashkin} 
\affiliation[label1]{organization={Skolkovo Institute of Science and Technology, Artificial Intelligence Center},
            addressline={Bolshoy Boulevard, 30, bld. 1}, 
            city={Moscow},
            postcode={121205}, 
            country={Russsia}}
\affiliation[label2]{organization={Digital Materials LLC},
            addressline={Kutuzovskaya str. 4A}, 
            city={Odintsovo},
            postcode={143001}, 
            state={Moscow region},
            country={Russsia}}
\author[label1]{Olga Klimanova}
\author[label1,label2]{Alexander Shapeev}


\begin{abstract}
Conflicting experiments disagree on whether the titanium–vanadium (Ti–V) binary alloy exhibits a body-centred cubic (BCC) miscibility gap or remains completely soluble.
A leading hypothesis attributes the miscibility gap to oxygen contamination during alloy preparation.
To resolve this disagreement, we use an \emph{ab initio}\,+\,machine-learning workflow that couples an actively-trained Moment Tensor Potential with Bayesian inference of free energy surface.
This workflow enables construction of the Ti–V phase diagram across the full composition range with systematically reduced statistical and finite-size errors.
The resulting diagram reproduces all experimental features, demonstrating the robustness of our approach, and clearly favors the variant with a BCC miscibility gap terminating at $T_\mathrm{crit}=980 \;\text{K}$ and $c_\mathrm{crit}=0.67$. 
Because our simulations model a perfectly oxygen-free Ti–V system, the observed gap cannot originate from impurity effects, in contrast to recent CALPHAD reassessments.
\end{abstract}

\begin{graphicalabstract}
\includegraphics[width=\textwidth]{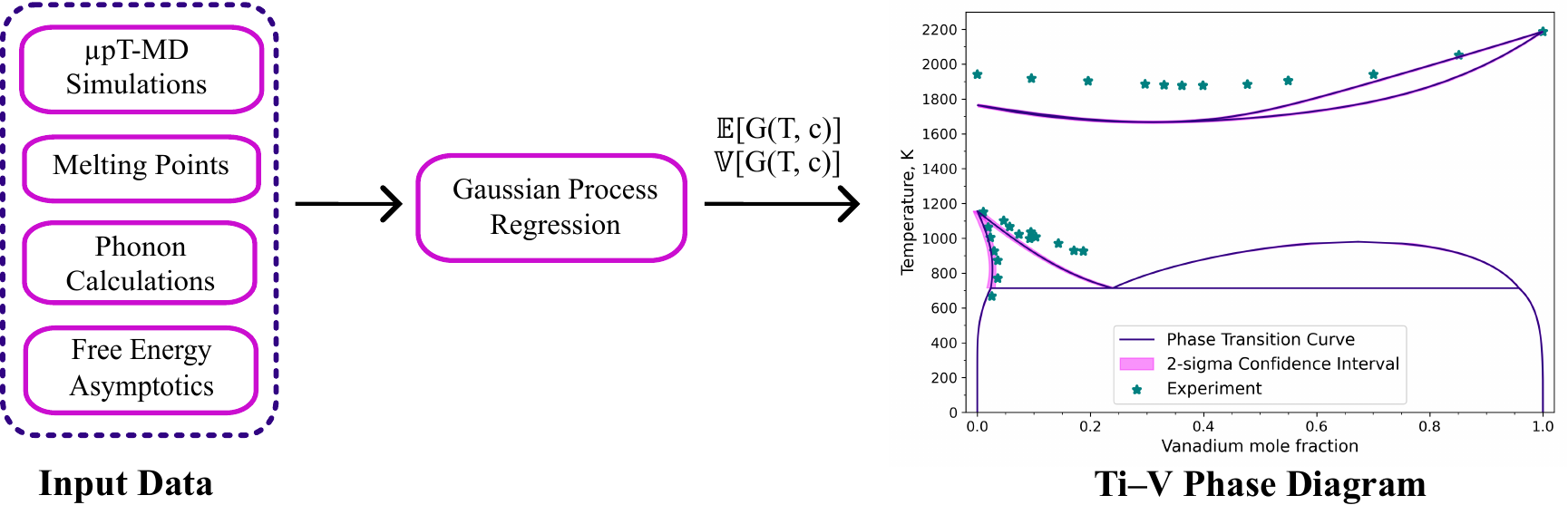}
\end{graphicalabstract}

\begin{highlights}

\item Ti–V phase diagram constructed with \emph{ab initio} accuracy using actively trained machine-learning interatomic potential. 

\item Statistical and finite-size errors of phase transition boundaries are systematically eliminated.

\item Computed Ti–V phase diagram reproduces all experimental features and indicates a body-centered cubic miscibility gap.

\item Findings challenge the earlier hypothesis attributing the miscibility gap to oxygen impurities. 

\end{highlights}


\end{frontmatter}



\section{Introduction}
Phase diagrams are foundational tools in materials science, serving as guides for designing and synthesizing materials under specific thermodynamic conditions. Recent advances in machine-learned interatomic potentials \cite{shapeev2016-mtp, bartok2010gap, bartok2015tutorial, drautz2019ace, batzner2022nequip, musaelian2023allegro, wang2018deepmd, zhang2023deepmdkitv2, batatia2022mace, yang2024mattersim} now enable molecular simulations with \emph{ab initio} accuracy at a fraction of the computational cost. Despite these significant advances, constructing phase diagrams from first principles remains challenging.

The most accurate method for phase diagram construction is thermodynamic integration \cite{Berend_Smit_MD, Kruglov_nitrogen, grabowski2019-free-energy-hea}, which reconstructs the free energy from molecular dynamics (MD) simulations by integrating its derivatives with respect to simulation parameters along a path to a reference state. While being very accurate, this method requires multiple MD simulations to discretize each integration path, making it computationally expensive to map the entire free energy surface. Another widely used approach is coexistence simulations \cite{silicon_phad_2018_bartok,interface_pinning_2013, Ga-As_phad_2021}, where phases are simulated in direct contact. However, because such simulations must be repeated for each set of conditions, they become impractical for systems with many phase transition boundaries.

A recently proposed Bayesian framework \cite{ladygin2021-phad, miryashkin2023} retains the strengths of free-energy reconstruction while avoiding the drawbacks of both approaches. MD simulations are performed only once for each selected temperature and composition, while systematically eliminating finite-size effects and propagating statistical uncertainties from the MD data into the predicted free energies and phase boundaries. This approach produces the full phase diagram while substantially reducing the number of required simulations.

\begin{figure}[h]
\centering
\includegraphics[width=0.8\textwidth]{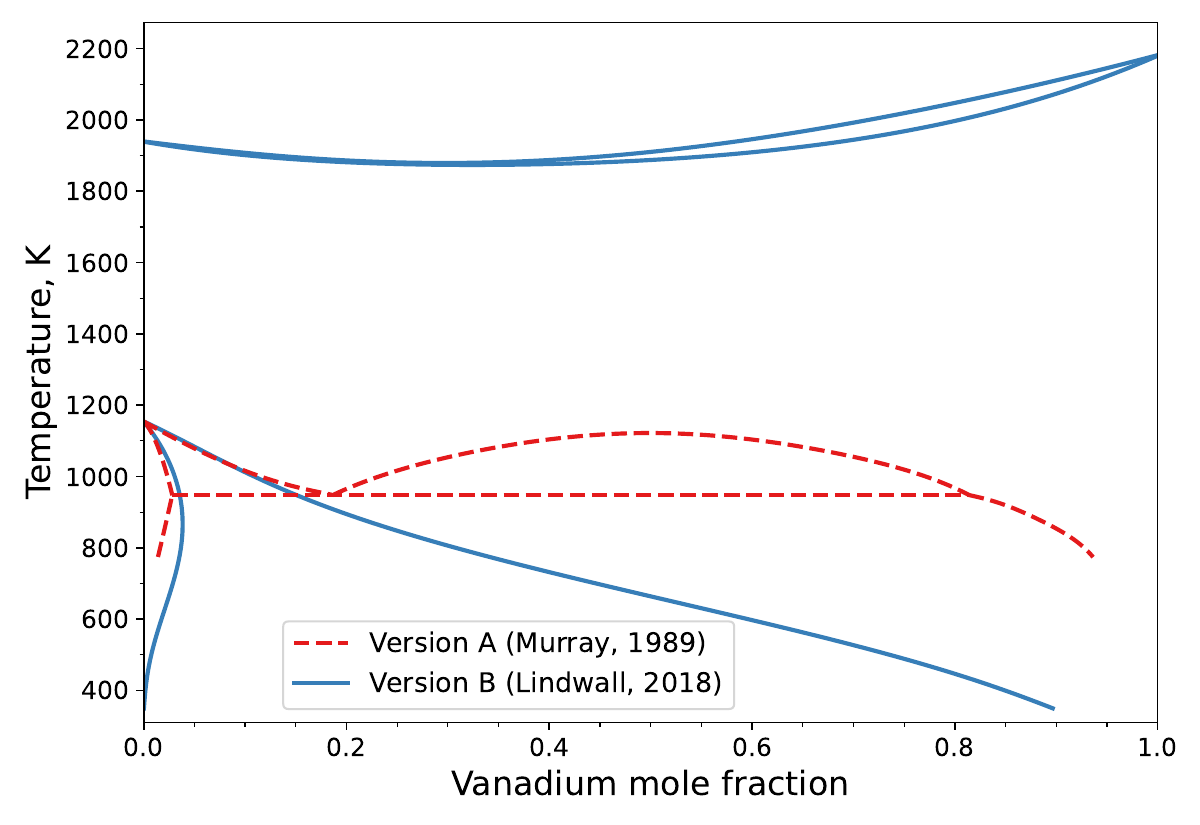}
\caption{
Comparison of two conflicting interpretations of the Ti–V binary phase diagram.
Version A predicts a BCC-phase miscibility gap, whereas Version B shows complete solid solubility across the entire composition range.
Version A is adapted from Murray (1989) \cite{murray1989phase}, and Version B is adapted from Lindwall et al. (2018) \cite{Lindwall2018}. 
}
\label{fig:compare_experiments}
\end{figure}

In this work, we apply the Bayesian framework to the titanium–vanadium (Ti–V) binary system, a case with long-standing experimental controversy. One set of studies \cite{nakano1980phase, murray1989phase} identifies a body-centred cubic (BCC) miscibility gap (Figure \ref{fig:compare_experiments}, version A), while others \cite{khaled1978phase, khaled1981effect, Fuming01121989} claim complete solid solubility across the entire composition range (Figure \ref{fig:compare_experiments}, version B). The most recent CALPHAD (CALculation of PHAse Diagrams)\cite{calphad_book_1998} reassessment by Lindwall \cite{Lindwall2018} also indicates complete solid solubility, attributing earlier reports of a miscibility gap to oxygen impurities. 

We infer the complete temperature–composition Ti-V phase diagram with confidence intervals for each phase boundary using the Bayesian framework.
The resulting diagram reproduces all experimental features, demonstrating the robustness of our approach, and clearly favors the variant with a BCC miscibility gap, showing that oxygen impurities, absent by construction in our computations, are unnecessary for gap formation.

Our workflow proceeds in three main stages:
\textbf{(i)} an active-learning protocol based on the D-optimality criterion selects \emph{ab initio} atomic configurations, from which we fit an interatomic potential; 
\textbf{(ii)} we employ the fitted potential to generate a thermodynamic data set — with associated computational uncertainties — for each equilibrium phase: the body-centred cubic (BCC) solid, the hexagonal-close-packed (HCP) solid, and the liquid;
\textbf{(iii)} from these data, we reconstruct free-energy surfaces and locate the corresponding phase-transition boundaries.
The workflow is readily transferable to other alloy systems — especially those where experimental information is sparse or conflicting.

The remainder of the paper is organized as follows. Section \ref{sec:related_work} reviews related work. Section \ref{sec:methods} presents the Bayesian algorithm for phase-diagram inference. Section \ref{subsec:mtp_potential} details the construction of the interatomic potential. Finally, Section \ref{subsec:tiv_phase_diag} describes the generation of the thermodynamic data set and the inference of the Ti–V phase diagram.

\section{Related Work}
\label{sec:related_work}

Two first-principles investigations have explored the Ti--V phase diagram, each adopting distinct approximations.
Sluiter \emph{et al.}\,\cite{Sluiter1991} combined a tight-binding Hamiltonian --- whose hopping integrals are fitted once to linear augmented-plane-wave (LAPW) band structures --- with the Coherent Potential Approximation (CPA) and the Generalized Perturbation Method (GPM), using the Cluster Variational Method (CVM) to compute configurational free energies.
More recently, Chinnappan \emph{et al.} \cite{CHINNAPPAN2016125} employed a cluster expansion parametrized on density functional theory calculations, combined with Monte Carlo sampling. To capture vibrational effects, they used the bond stiffness versus bond length method, which assumes a linear relationship between bond length and the nearest-neighbour force constant.

The approaches outlined above are limited by the fidelity of the underlying energetics, which restricts the precision of their phase-boundary predictions. 
Modern machine-learning interatomic potentials, such as MTPs \cite{shapeev2016-mtp, novikov2020-mlip}, deliver near \emph{ab initio} accuracy for both energies and forces while running several orders of magnitude faster than conventional \emph{ab initio} methods. 
Moreover, anharmonic vibrational contributions can be evaluated more rigorously through molecular-dynamics simulations driven by such high-fidelity interatomic potentials.

The semi-empirical CALPHAD approach has also been applied to the Ti--V system.
Model parameters are fitted to experimental data (e.g., calorimetric and phase-equilibrium measurements) via least-squares optimization.  Yang \emph{et al.}\,\cite{YANG2017365} and Lindwall \emph{et al.}\,\cite{Lindwall2018} proposed updated CALPHAD descriptions: the former predicts a miscibility gap in the BCC phase, whereas the latter does not. Because CALPHAD relies on experimental datasets that may be sparse or mutually inconsistent, its predictive power is inherently limited for systems — such as Ti--V — where contradictory measurements exist.

We also note the recent experimental studies of Smith et al. \cite{Smith_2017} and MacLeod et al. \cite{MacLeod_2021} on commercial titanium alloys (e.g., Ti-6Al-4V, Ti-36Nb-2Ta-0.3O) that investigate the effects of oxygen impurities on phase transition boundaries and thermodynamic properties of these alloys.

\section{Methods}  
\label{sec:methods}
\begin{figure*}
    \includegraphics[width=15cm]{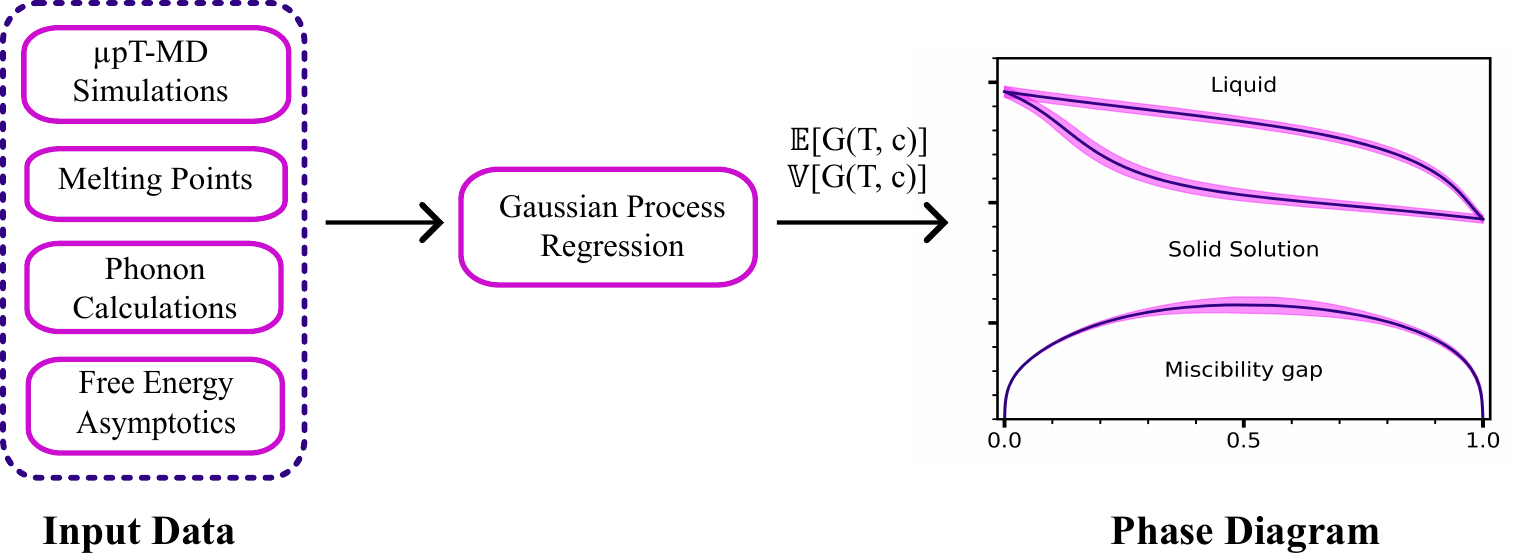}
    \caption{Schematic illustration of our algorithm. Thermodynamic data of different natures (MD simulations, melting points, phonon computations) along with the asymptotics of the free energies is fed as input data to the Gaussian process regression. We next predict the free energies with their uncertainties to construct the phase diagram.
    }
    \label{fig:algo_scheme}
\end{figure*}

This section outlines the Bayesian learning framework to infer phase diagrams. The algorithm features the propagation of uncertainties in heterogeneous thermodynamic data to the resulting phase diagram (Fig. \ref{fig:algo_scheme}). 

The core of the algorithm is a Gaussian process regression that combines heterogeneous data sources—molecular-dynamics (MD) simulations, melting points, and phonon computations—along with their associated uncertainties. 
Molecular-dynamics simulations are carried out in the semi-grand canonical $\mu P T$ ensemble at zero pressure.
From the learned free energies of each phase we reconstruct phase-transition boundaries in the thermodynamic limit \(N \rightarrow \infty\) together with  confidence intervals for every boundary.  
A full description of the algorithmic details is given in Ref.~\cite{miryashkin2023}.

Rather than modeling the Gibbs free energy \(G(T,c,N)\) directly, we recast the target quantity as
\begin{align}
   G(T,c,N) \;=\; G_{\text{ref}}(T,c,N) \;-\; T\,S(T,c,N),
   \label{eq:S_def}
\end{align}
where \(G_{\text{ref}}\) is a suitably chosen reference free energy, \(T\) is the temperature (reported in electron-volts, adopting \(k_B=1\)), \(c\) is the atomic concentration, and \(N\) is the total number of atoms in the system. If $G_{\text{ref}}(T, c, N)$ is chosen as $G(0, c, N)$ then $S(T, c, N)$ is the conventional entropy. If $G_{\text{ref}}(T, c, N)$ is chosen to be the free energy of an ideal gas, then the resulting $S(T, c, N)$ would typically be called the excess entropy. By analogy, we refer to $S(T, c, N)$ as \emph{entropy}, although our $G_{\text{ref}}(T, c, N)$ will be somewhat different from the common ones and will be chosen to facilitate the fitting of $S(T, c, N)$. Learning \(S\) sidesteps the singularities inherent to \(G\) (like $c \log c$), yielding smoother fits and reliable uncertainty estimates.

The remainder of this section is organized as follows. Section \ref{sec:gaus_pross_regression} introduces the Gaussian process regression framework. The subsequent subsections describe how different data sources are incorporated into the algorithm: free-energy asymptotics (Section \ref{subsec:free_energy_asympt}), molecular dynamics data (Section \ref{subsec:molecular_dyn_data}), phonon calculations (Section \ref{subsec:phonon_calculations}), and melting points (Section \ref{subsec:melt_points}). Section \ref{subsec:coexistence_algo} focuses specifically on the Bayesian estimation of the melting temperature and its associated confidence interval. Finally, Section \ref{subsec:mtp} presents the functional form of the moment-tensor potential employed in this study.

\subsection{Gaussian Process Regression}
\label{sec:gaus_pross_regression}
We reconstruct the entropy of each phase using Gaussian process regression (GPR) \cite{rasmussen2006-book-gaussian}, which naturally incorporates the uncertainty in the input data. As the simulation data is approximately normally distributed, the Gaussian process provides an appropriate probabilistic model for the entropy.

In GPR, the entropy \( S(T, c, N) \) is modeled as a Gaussian process, where the covariance between two input points \((T_1, c_1, N_1)\) and \((T_2, c_2, N_2)\) is defined via a kernel function:
\[
	\begin{array}{r}
k[(T_1, c_1, N_1), (T_2, c_2, N_2)]
\hspace{12em}\mathstrut
\\
:= {\rm Cov}[S(T_1, c_1, N_1), S(T_2, c_2, N_2)].
\end{array}
\]
Assuming that the entropy is a smooth function of \(T\), \(c\), and \(N^{-1}\), we adopt a squared exponential kernel. For crystalline phases, we use:
\begin{align}\label{eq:cryst_ker}
\begin{split}
    k_{\rm cryst}&:=  \theta_0^2 + \theta_f^2  
    \exp \left( -\frac{(T_1 - T_2)^2}{2 \theta_T^2}\right) 
    \exp \left( -\frac{(c_1 - c_2)^2}{2 \theta_c^2} \right) 
    \\&\phantom{:=\mathstrut} 
    \cdot \exp  \left( -\left(\frac{1}{N_1} - \frac{1}{N_2} \right)^2 \frac{\theta_N^2}{2} \right),
\end{split}
\end{align}
where \(\theta_i\) are learnable hyperparameters. Notably, this kernel is constructed to remain well-defined in the limit \(N \to \infty\), enabling extrapolation to the thermodynamic limit based on simulations at finite sizes.

For the liquid it is important to explicitly take into account that
the energy is defined up to two additive constants (an arbitrary energy shift for each of the species).  Therefore we allow the entropy to have a linear in $c$ shift, $S \sim \theta_1 c/T + \theta_2 (1-c)/T$, which features as the last two terms in our kernel for the liquid phase:
\begin{align*}
    k_{\rm liq} &:=  \theta_0^2 + \theta_f^2  
    \exp \left( -\frac{(T_1 - T_2)^2}{2 \theta_T^2}\right) 
    \exp \left( -\frac{(c_1 - c_2)^2}{2 \theta_c^2} \right)  
\\ &\mathstrut\phantom{:=\mathstrut}\mathstrut
    \cdot \exp  \left( -\left(\frac{1}{N_1} - \frac{1}{N_2} \right)^2 \frac{\theta_N^2}{2} \right)
\\ &\mathstrut\phantom{:=}\mathstrut
	+ \theta_{1}^2 \frac{c_1 c_2}{T_1 T_2} 
    + \theta_{2}^2 \frac{(1-c_1) (1-c_2)}{T_1 T_2}.
\end{align*}

A key characteristics of Gaussian processes is that any linear functional of a Gaussian process is also normally distributed. This is essential in our case, as the input data include not only entropy values but also their derivatives. For instance, the covariance between the entropy and its temperature derivative is given by:
\[
{\rm Cov}\left[\frac{\partial }{\partial T_1} S(T_1, c_1), S(T_2, c_2)\right] = \frac{\partial }{\partial T_1} k[(T_1, c_1), (T_2, c_2)].
\]
In general, any data point can be viewed as a linear functional \(X\) acting on \(S\), such as \( \<S | X\> = S(T, c) \) or \( \<S | X\> = \frac{\partial S}{\partial T}(T, c) \). Accordingly, we generalize the kernel definition to arbitrary functionals $X_1$ and $X_2$:
\[
k(X_1, X_2) := {\rm Cov}[\< S| X_1 \>, \< S| X_2 \>].
\]

We now discuss the Gaussian process regression algorithm in which we use the kernels defined above.
We perform calculations in the data point (functional) $X_i$ for which we obtain the target value $Y_i$ with the Gaussian noise $\Delta Y_i$, and hence require that the input point has the Gaussian distribution:
\[
\< S | X_i\> \sim \mathcal{N} \big( Y_i, (\Delta Y_i)^2 \big).    
\]
Thus, the input data to the regression algorithm has the form of tuples $(X_i, Y_i, \Delta Y_i)$. 
To predict the value \(Y_* = \<S | X_*\>\) at a new point \(X_*\), we form a joint Gaussian distribution over \(\{Y_i\}\) and \(Y_*\), with zero mean and covariance:
\[
	{\rm cov}
	\begin{pmatrix}
		\boldsymbol{Y} \\
		Y_*
	\end{pmatrix}
=
	\begin{pmatrix}
		K(\boldsymbol{X}, \boldsymbol{X})+\operatorname{diag}(\boldsymbol{\Delta} \boldsymbol{Y}^2) & K\left(\boldsymbol{X}, X_*\right) \\
		K\left(X_*, \boldsymbol{X}\right) & K\left(X_*, X_*\right)
	\end{pmatrix},
\]
where ${\bm X}$ and ${\bm Y}$ are the vectors composed of $X_i$ and $Y_i$, $\operatorname{diag}(\boldsymbol{\Delta Y}^2)$ is the matrix with $(\Delta Y_i)^2$ on the diagonal, and  $K(\bm{X}, \bm{X})$ is the matrix composed of $k(X_i, X_j)$. 

Following \cite{bishop2006pattern}, the posterior distribution of \(Y_*\) is Gaussian with:
\begin{align} \notag
\mathbb{E}[Y_*] &= K(X_*, \bm{X}) K_y^{-1} \bm{Y}, \qquad \text{and}
\\ \label{eq:var_gp}
\mathbb{V}[Y_*] &= K(X_*, X_*) - K(X_*, \bm{X}) K_y^{-1} K(\bm{X}, X_*),
\end{align}
where \(K_y = K(\bm{X}, \bm{X}) + \operatorname{diag}(\bm{\Delta Y}^2)\).

Gaussian process regression is a non-parametric machine learning method that is defined by its kernel and a set of hyperparameters, such as \({\bm \theta} = (\theta_0, \theta_f, \theta_T, \theta_c, \theta_N)\) in Eq.~\eqref{eq:cryst_ker}. These hyperparameters are learned by maximizing the marginal likelihood \(p({\bm Y} \mid {\bm X}, {\bm \theta})\), which represents the probability of observing the targets \({\bm Y}\) given the inputs \({\bm X}\) and hyperparameters \({\bm \theta}\). The log marginal likelihood is given by
\begin{align*}
    \log p({\bm Y} \mid {\bm X}, {\bm \theta}) 
    &= -\tfrac{1}{2} {\bm Y}^{\mathsf{T}} K_y^{-1} {\bm Y}
    - \tfrac{1}{2} \log |K_y|
    - \tfrac{n}{2} \log(2\pi),
\end{align*}
where \(n\) is the number of input points.

\subsection{Free energy asymptotics}
\label{subsec:free_energy_asympt}

In this subsection, we show how we select $G_{\text{ref}}(T, c, N)$ for the crystalline and liquid phases.

For the crystalline phases, the free energy as $T \rightarrow 0$ has the form 
\begin{align}\label{eq:f_cryst}
\begin{split}
	G^{\rm cryst} (T, c, N)
	&= 
	E_0
    + T c \log(c)
    + T (1 - c) \log(1 - c) \\
	&- T \log(N) + T - {\textstyle \frac32} T \log(2\pi T) 
    \\&
    + {\textstyle \frac12} T N^{-1} \log\det \hat{H}_0(N)
	\\&
	 + O\big(T\big)o(1) + O(T^2)
	 ,
\end{split}
\end{align}
where $E_0$ is the energy of the ground state structure, $H_0$ is the energy Hessian at the ground state and $o(1)$ denotes a vanishing as $N\to\infty$ term. 
Note that $ \log\det \hat{H}_0(N)$ scales as $O(N)$, therefore $N^{-1} \log\det \hat{H}_0(N)$ is a nonvanishing intensive quantity.
The formula above explicitly accounts for the vibrational and configurational contributions to the free energy.
We select the following reference free energy for the crystalline phase:
\begin{align*}
\begin{split}
G^{\rm cryst}_{\rm ref}(T, c, N)
&:= 
E_0(c)
+ T c \log(c)
+ T (1-c) \log(1-c) \\
&- T \log(N) + T - {\textstyle \frac32} T \log(2\pi T),
\end{split}
\end{align*}
where $E_0(c)$ is the linear interpolation between the energies of monoatomic ground-state structures.

For the liquid phase, we select the reference free energy that accounts only for the configurational contribution 
\begin{align*}
\begin{split}
G^{\rm liq}_{\rm ref}(T, c, N)
&:= 
T c \log(c) + T (1-c) \log(1-c) 
\\&- T \log(N).
\end{split}
\end{align*}

\subsection{Molecular Dynamics Data}
\label{subsec:molecular_dyn_data}
Thermodynamic observables are extracted from molecular-dynamics (MD) simulations carried out in the semi-grand-canonical $\mu PT$ ensemble, where $\mu$ denotes the chemical potential, $P$ the external pressure, and $T$ the temperature.
Ensemble averages are denoted by angle brackets, e.g.\ the mean concentration $c\equiv\langle\chi\rangle$ and the mean energy $\langle E\rangle$.

In this ensemble the derivatives of the Gibbs free energy can be written directly in terms of these averages: 
\begin{align}
    \frac{\partial G(T, \<\chi\>)}{\partial \<\chi\>} &= \mu,
    \label{eq:dF_dc}
    \\
    \frac{\partial (\beta G(T, \<\chi\>))}{\partial T} &= -\frac{\< E \>}{T^2},
    \label{eq:dF_dT}
\end{align}
with $\beta\equiv 1/T$.

Combining Eqs.\,\eqref{eq:dF_dc}–\eqref{eq:dF_dT} with the definition of the entropy, Eq.\,\eqref{eq:S_def}, gives
\begin{align*}
	\frac{\partial S}{\partial T} &= \frac{\<E\>}{T^2} + \frac{\partial (\beta G^{\rm ref})}{\partial T},
	\\ \notag
	\frac{\partial S}{\partial c} &= - \beta \mu + \frac{\partial (\beta G^{\rm ref})}{\partial c}.
\end{align*}

The Gaussian-process regression incorporates these derivative data together with their statistical uncertainties—arising from the finite length of each MD trajectory—so that every observation weighted according to its precision. 

\subsection{Phonon Calculations}
\label{subsec:phonon_calculations}

As in thermodynamic integration, the $\mu PT$ simulations provide the temperature and composition derivatives of the entropy \(S(T,c,N)\).
These derivatives are sufficient to reconstruct \(S\) for each phase up to an additive constant; however, accurate phase‐boundary predictions require that constant to be fixed as well.

For solid phases we determine the constant from the harmonic limit,
\begin{align}\label{eq:ref_solid}
\begin{split}
S(0,c,N) &=
\lim_{T\to0} T^{-1} (G_{\rm ref}(T,c,N) - G(T,c,N))
\\&=
-{\textstyle \frac12} N^{-1} \log\det \hat{H}_0(N).
\end{split}
\end{align}
where \(\hat{H}_0(N)\) is the \(3N\times 3N\) dynamical (Hessian) matrix evaluated at zero temperature.

For the liquid phase the additive constant is determined from the melting temperature obtained via \(NPT\) solid–liquid coexistence simulations, as described in Sec. \ref{subsec:melt_points}.


\subsection{Melting Points}
\label{subsec:melt_points}

Thanks to the Bayesian approach, our algorithm incorporates the data on melting points along with their corresponding uncertainties. Melting temperatures are evaluated by the coexistence algorithm summarized in Sec. \ref{subsec:coexistence_algo}, following the procedure of Ref.~\cite{Klimanova_2023}.

Each melting point enters the algorithm through the linear functional
\[
G_2 (T_{\rm m}, c_{\rm m}, N) - G_1 (T_{\rm m}, c_{\rm m}, N) = 0,
\] 
where the subscripts 1 and 2 denote the two coexisting phases (e.g.\ solid and liquid).  
The corresponding uncertainty in the linear functional is given by
\[
\Delta\left( G_2(T_{\rm m}, c_{\rm m}, N) - G_1(T_{\rm m}, c_{\rm m}, N) \right) = \left| \frac{\partial G_2}{\partial T} - \frac{\partial G_1}{\partial T} \right| \Delta T_{\rm m}.
\]
so that the uncertainty in the measured melting temperature, $\Delta T_{\mathrm m}$, is transferred to the free energy difference and ultimately to the inferred phase diagram.

\subsection{Bayesian Coexistence Algorithm for Melting Point Prediction}
\label{subsec:coexistence_algo}

Our melting-point calculations follow the autonomous, physics-informed workflow of Klimanova \textit{et al.}\,\cite{Klimanova_2023}.
The workflow is organized in two hierarchical levels:  
on the first level we infer the melting temperature \(T^{*}\) for a fixed number of atoms \(N\),  
whereas on the second level we model the dependence of \(T^{*}\) on \(N\) and extrapolate it to the thermodynamic limit \(N \to \infty\).

\paragraph{Level 1: Bayesian inference at fixed system size.}
We run solid–liquid coexistence simulations in the isothermal–isobaric (NPT) ensemble.  
For each chosen \(N\) the output is a data set  
$(T^{(i)}, n_{\rm s}^{(i)}, n_{\rm l}^{(i)})_{i=1,\ldots, n}$,  
where \(T^{(i)}\) is the molecular-dynamics temperature, and  
$n_{\rm s}^{(i)}$ and $n_{\rm l}^{(i)}$ are the numbers of “solid’’ and “liquid’’ outcomes, respectively, defined by the final state of the trajectory.

We then apply non-linear Bayesian regression to these data.  
Knowing the likelihood of observing a given set of outcomes as a function of the trial temperature \(T\) and its spread \(\sigma\) ~\eqref{eq:data_prob_density},  
we can ask — with Bayes’ theorem \eqref{eq:bayes} — how probable it is that \(T\) and \(\sigma\) are indeed the melting temperature of the system and its uncertainty:

\begin{equation}\label{eq:data_prob_density}
    p(\text{data}\vert T, \sigma) = \frac{\prod\limits_{i = 1}^n \exp \left({n_{\rm l}^{(i)}\cdot\frac{T^{(i)} - T}{\sigma}}\right)}{\prod\limits_{i = 1}^n \big(1 + \exp\left({\frac{T^{(i)} - T}{\sigma}}\big)\right)^{n_{\rm l}^{(i)} + n_{\rm s}^{(i)}}},
\end{equation}

\begin{equation}\label{eq:bayes}
    p(T, \sigma\vert \text{data}) = \frac{p(\text{data}\vert T, \sigma)\,p(T)\,p(\sigma)}{p(\text{data})}.
\end{equation} 


In Eq.~\eqref{eq:bayes} we take the evidence \(p(\text{data})\) and the prior \(p(T)\) to be uniform,  
while for the spread we adopt the scale-invariant prior \(p(\sigma)=\sigma^{-2}\), favoring small \(\sigma\).
    
The posterior \eqref{eq:bayes} yields the mean melting temperature and its variance,

\begin{align} \label{eq:T_melt}
T^* &= \overline{T} =  \frac{\iint T\cdot p(T,\sigma\vert \text{data})\,{\rm d}T\,\sigma^{-2}\,{\rm d}\sigma}{\iint p(T,\sigma\vert \text{data})\,{\rm d}T \,\sigma^{-2}\,{\rm d}\sigma},
\\ \label{eq:var_Tmelt}
(\Delta T^*)^2 &= 
{\rm Var}(T) = 
\frac{\iint (T- T^*)^2\cdot p(T,\sigma\vert \text{data})\,{\rm d}T \, \sigma^{-2} \, {\rm d}\sigma}{\iint p(T,\sigma\vert \text{data}) \, {\rm d} T \, \sigma^{-2} \, {\rm d}\sigma}.
\end{align}

Thus, after level 1 we obtain triplets $(T^*_i (N), \Delta T^*_i, N_i)$,  
where \(T^{*}_{i}\) is the mean melting temperature and \(\Delta T^{*}_{i}\) its uncertainty for a simulation cell containing \(N_{i}\) atoms.

\paragraph{Level 2: Extrapolation to the thermodynamic limit.}
To link the finite-size values \(T^{*}(N)\) to the bulk melting point we employ Gaussian-process regression with the kernel

\begin{equation}\label{eq:kernel}
    k(N_{1},N_{2})=
    \theta_{f}^{2}\,
    \exp\!\Bigl[-\bigl(N_{1}^{-1}-N_{2}^{-1}\bigr)^{2}\,\theta_{N}^{2}/2\Bigr],
\end{equation}
where the hyper-parameters \(\theta_{f}\) and \(\theta_{N}\) are optimized during training.  
The kernel embodies the expectation that covariances decay as \(\mathcal{O}(1/N)\).  
Consequently it correlates melting temperatures for finite system sizes with its value at \(N\to\infty\),  
allowing us to predict the melting temperature together with its uncertainty in the thermodynamic limit.

\subsection{Moment Tensor Potential}
\label{subsec:mtp}

In this work, we employ the Moment Tensor Potential (MTP) as implemented in the MLIP-2 package\cite{novikov2020-mlip}. Within the MTP framework, the total potential energy of an atomic system is expressed as the sum of local contributions from each atom:

\[
E^{\mathrm{MTP}} = \sum_{i=1}^{N} V(\mathbf{n}_i),
\]

where the index \( i \) runs over all \( N \) atoms in the system, and \( \mathbf{n}_i \) denotes the local atomic neighborhood of atom \( i \) within a cutoff radius \( R_\mathrm{cut} \). The local energy \( V(\mathbf{n}_i) \) is modeled as a linear combination of basis functions:

\[
V(\mathbf{n}_i) = \sum_\alpha \xi_\alpha B_\alpha(\mathbf{n}_i),
\]

where \( \xi_\alpha \) are the coefficients to be fitted, and \( B_\alpha(\mathbf{n}_i) \) are basis functions dependent on the atomic environment.

To ensure the descriptors are invariant under translation, rotation, and permutation of atoms, the MTP framework introduces moment tensor descriptors \( M_{\mu,\nu}(\mathbf{n}_i) \), defined as:

\[
M_{\mu,\nu}(\mathbf{n}_i) = \sum_j f_\mu(|\mathbf{r}_{ij}|, z_i, z_j)\, \underbrace{\mathbf{r}_{ij} \otimes \cdots \otimes \mathbf{r}_{ij}}_{\nu \text{ times}},
\]

where the sum is taken over all neighbors \( j \) of atom \( i \), \( \mathbf{r}_{ij} \) is the vector from atom \( i \) to atom \( j \), and \( z_i, z_j \) denote the atomic species of atoms \( i \) and \( j \), respectively. The operator \( \otimes \) denotes the outer product of vectors, yielding a rank-\( \nu \) tensor. The function \( f_\mu \) encodes the radial dependence of the moment tensor and is given by:

\[
f_\mu(|\mathbf{r}_{ij}|, z_i, z_j) = \sum_\beta c^{(\beta)}_{\mu, z_i, z_j} Q^{(\beta)}(|\mathbf{r}_{ij}|),
\]

where \( c^{(\beta)}_{\mu, z_i, z_j} \) are trainable parameters and \( Q^{(\beta)} \) are radial basis functions defined as:

\[
Q^{(\beta)}(|\mathbf{r}_{ij}|) = T^{(\beta)}(|\mathbf{r}_{ij}|) \left(R_\mathrm{cut} - |\mathbf{r}_{ij}|\right)^2.
\]

Here, \( T^{(\beta)}(|\mathbf{r}_{ij}|) \) denotes the Chebyshev polynomial of degree \( \beta \), defined on the interval \( [R_\mathrm{min}, R_\mathrm{cut}] \), and the term \( (R_\mathrm{cut} - |\mathbf{r}_{ij}|)^2 \) ensures a smooth decay to zero at the cutoff boundary.

The moment tensor descriptors \( M_{\mu,\nu} \) for \( \nu = 0, 1, 2, \ldots \) are tensors of varying rank, and the basis functions \( B_\alpha(\mathbf{n}_i) \) are constructed by contracting these tensors into rotationally invariant scalar quantities.
The number of possible contractions that yield scalar invariants is infinite. To control the number of basis functions used in practice, MTP introduces a descriptor level for moment tensors:
\[
\mathrm{lev}(M_{\mu,\nu}) = 2 + 4\mu + \nu.
\]

If a basis function \( B_\alpha \) is formed by contracting a set of moment tensors \( M_{\mu_1,\nu_1}, M_{\mu_2,\nu_2}, \ldots \), then its level is defined as:

\[
\mathrm{lev}(B_\alpha) = (2 + 4\mu_1 + \nu_1) + (2 + 4\mu_2 + \nu_2) + \cdots.
\]

By restricting the set of basis functions to those with \( \mathrm{lev}(B_\alpha) \leq d \), one obtains an MTP model of level \( d \) containing a finite number of basis terms. 

\section{Results}

\subsection{Development of the Moment Tensor Potential}
\label{subsec:mtp_potential}


We trained the machine-learning potential on configurations obtained from density-functional-theory (DFT) calculations performed with the Vienna Ab initio Simulation Package (VASP) \cite{PhysRevB.54.11169} using the projector-augmented-wave (PAW) method \cite{PhysRevB.59.1758}.
The Perdew–Burke–Ernzerhof generalized-gradient approximation (PBE-GGA) \cite{PhysRevLett.77.3865} was employed as the exchange–correlation functional. 
The self-consistent cycle was converged to an electronic-energy tolerance of $3 \times 10^{-5}$ eV with a Gaussian smearing width of 0.2 eV.
The \texttt{ENCUT} parameter was chosen as 466 eV which is $1.7\cdot\max(\texttt{ENMAX\_TI}, \texttt{ENMAX\_V})$, where \texttt{ENMAX} is the energy cutoff of the corresponding pseudopotential, and a $\Gamma$-centered k-point mesh having a spacing of 0.16 Å$^{-1}$ was used to ensure convergence of DFT computations up to 1 meV/at.

\begin{figure}[h]
    \centering
    \begin{subfigure}[t]{0.49\textwidth}
        \centering
        \includegraphics[width=\linewidth]{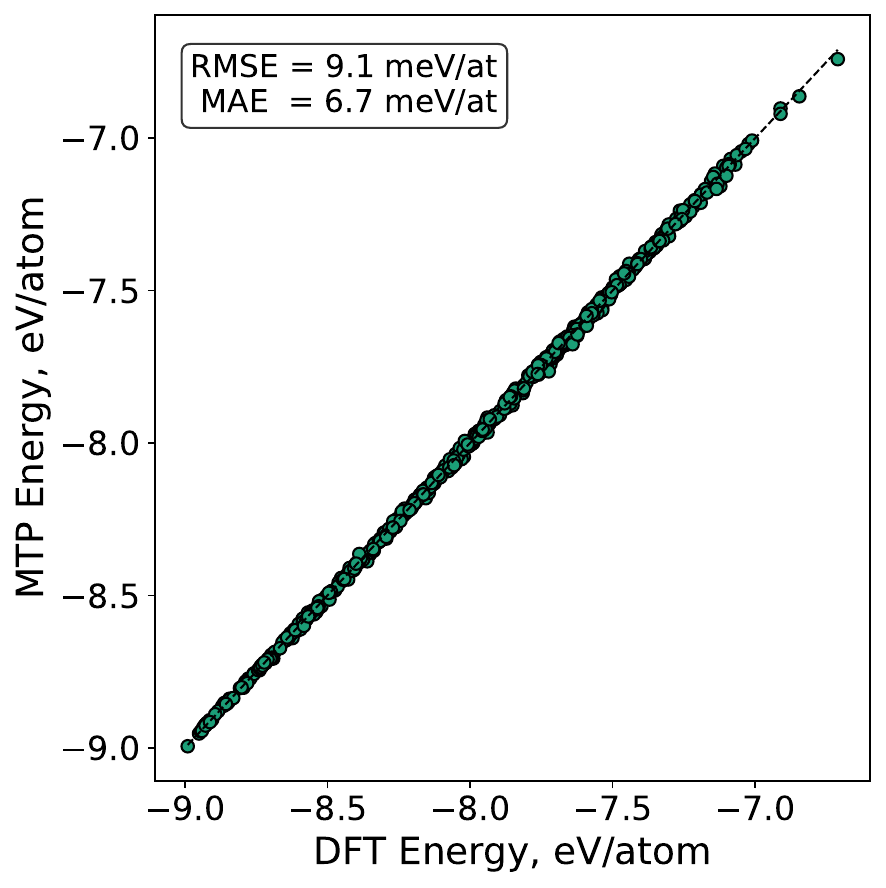}
        \caption{Comparison of MTP and DFT energies per atom.}
        \label{fig:sub_energy}
    \end{subfigure}
    \hfill
    \begin{subfigure}[t]{0.49\textwidth}
        \centering
        \includegraphics[width=\linewidth]{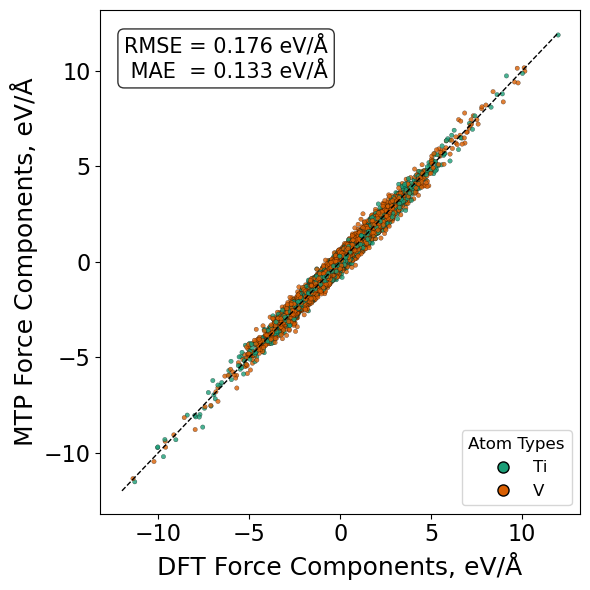}
        \caption{Comparison of MTP and DFT force components, color-coded by atom type.}
        \label{fig:sub_force}
    \end{subfigure}
    \caption{Evaluation of the MTP against DFT on the training dataset. A dashed black line represents a perfect linear dependence.}
    \label{fig:energy_force_comparison}
\end{figure}

In this study, we employed an active-learning strategy to train the interatomic potential, which has been shown to reduce markedly the number of required DFT calculations for constructing robust and accurate potentials \cite{PhysRevB.99.064114}. We adopted the D-optimality active-learning scheme \cite{PODRYABINKIN2017171} as implemented in the MLIP-2 package \cite{novikov2020-mlip}. A level-16 MTP served as the machine-learning interatomic potential.
The initial training set was generated from ab initio molecular-dynamics simulations. We then ran additional molecular-dynamics simulations at various chemical potentials and temperatures spanning the entire phase diagram, ultimately selecting 711 configurations for the training dataset.
The accuracy of the fitted potential on this dataset is illustrated in Figure \ref{fig:energy_force_comparison}; the root-mean-square error is $9.1\ \text{meV/atom}$ for energies and $176\ \text{meV}/\text{\AA}$ for forces.

\subsection{Phase Diagram Workflow}
\label{subsec:tiv_phase_diag}

\subsubsection{Exploratory sampling and surrogate model}

Having constructed the interatomic interaction potential based on DFT data in the previous section, we proceeded to collecting data from molecular dynamics simulations for the HCP, BCC, and liquid phases using the grand-canonical ensemble. We first studied the dependence of the concentration $c = c(T, \mu, N)$ using a supplementary dataset of points randomly sampled within the following ranges: $T \in [0,\text{K}, 3000,\text{K}]$, $\mu \in [-10, 10]$, and $N \in \{432, 686, 1024, 1458, 2000\}$ for the BCC and liquid phases, and $N \in \{216, 512, 1000, 1728\}$ for the HCP phase. We used short MD trajectories for this supplementary dataset, as our goal was to approximately estimate $c = c(\mu, T, N)$. 

Using the supplementary dataset, we trained a gradient boosting regressor (Scikit-learn implementation \cite{scikit-learn}) to capture the dependence of the concentration $c$ on $\mu$, $T$, and $N$. 
The resulting surrogate model enabled an almost uniform sampling of the 
$T$–$c$–$N$ space, yielding 62 data points in BCC region, 28 in HCP, and 36 in the liquid phase (Figure \ref{fig:tiv_whole_dataset}); these constitute the points for the MD simulations described below.

\subsubsection{Thermodynamic Dataset}

For each selected point, we performed MD simulations that were ten times longer than those in the supplementary dataset. Specifically, we ran eight independent trajectories, each consisting of 20,000 time steps with a time step of 1fs. We discarded points that underwent a phase transition during the simulation, which we identified using the polyhedral template matching algorithm \cite{Larsen_2016}. 

\begin{figure}[h]
    \centering
    \begin{subfigure}[t]{0.48\textwidth}
        \centering
        \includegraphics[width=\linewidth]{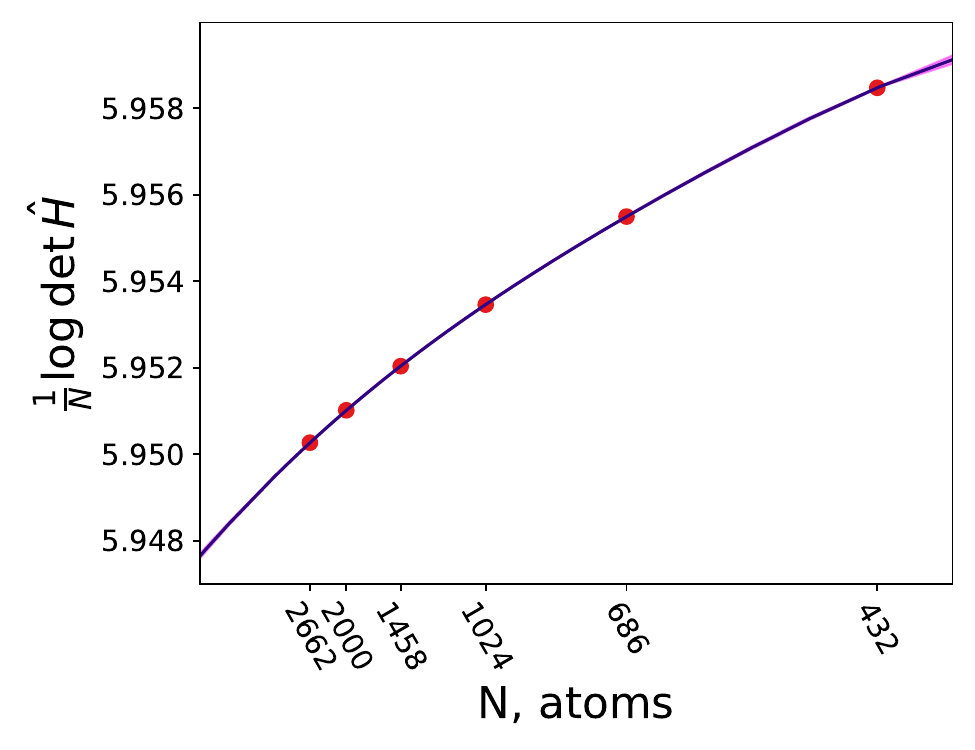}
        \caption{}
        \label{fig:hcp_ti_fit_log_det}
    \end{subfigure}
    \hfill
    \begin{subfigure}[t]{0.48\textwidth}
        \centering
        \includegraphics[width=\linewidth]{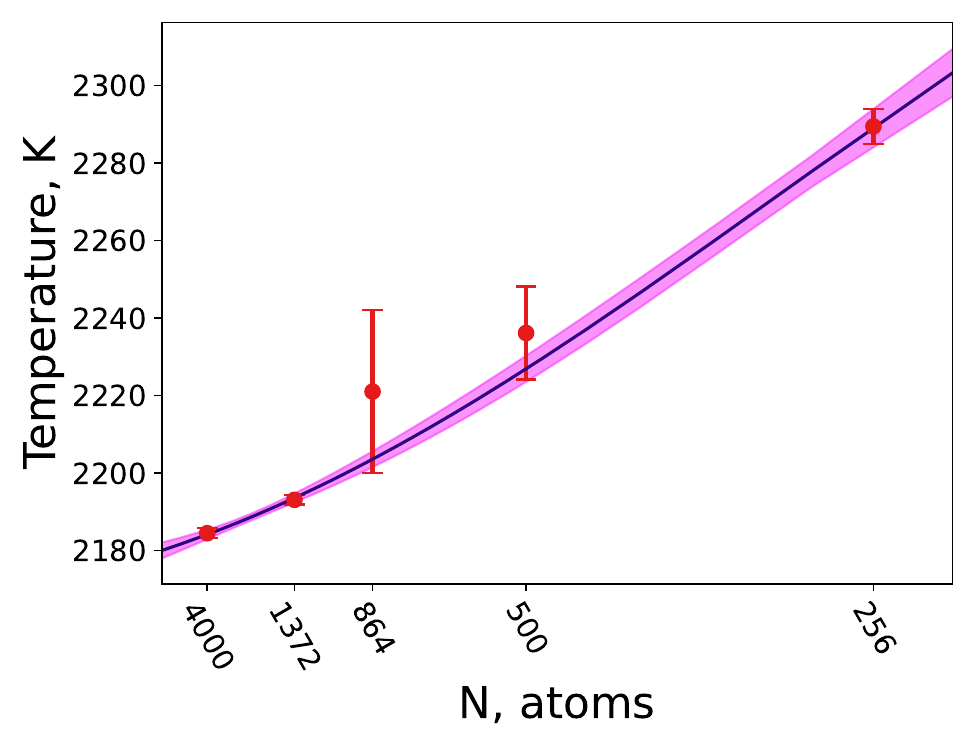}
        \caption{}
        \label{fig:bcc_v_melting_point}
    \end{subfigure}
    \caption{Dependence of $\frac{1}{N} \log\det \hat{H}$ (left panel), obtained from phonon computations, and   the melting point (right panel) for BCC vanadium on the number of atoms in the simulation cell. Gaussian process regression is used for extrapolation, yielding the following estimates in the thermodynamic limit ($N \rightarrow \infty$): $\frac{1}{N} \log\det \hat{H} = 5.94765 \pm 6\times10^{-5}$ and $T_{\mathrm{m}}(\mathrm{V}) = 2180 \pm 2 K$.}
\end{figure}

\begin{figure}[h]
\includegraphics[width=1.0\textwidth]{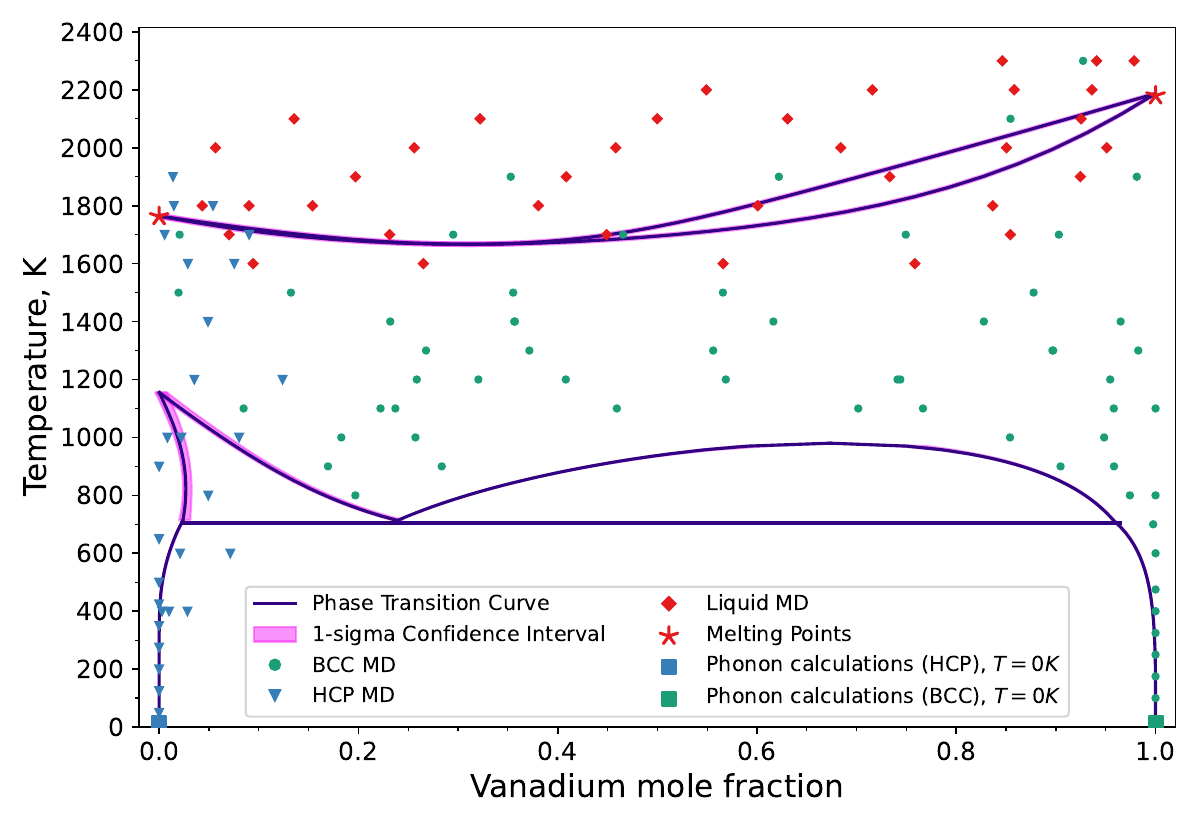}
\caption{
Dataset used to train the Bayesian algorithm, plotted atop the Ti–V phase diagram to illustrate the phase-stability domains.
The dataset comprises grand-canonical MD simulations, phonon calculations at 0 K for the solid phases, and melting points obtained from coexistence simulations.
Green markers denote the data used to reconstruct the HCP free-energy surface (MD simulations + phonon calculations), blue markers do the same for the BCC phase, and red markers (MD simulations + melting points) pertain to the liquid phase.   
}
\label{fig:tiv_whole_dataset}
\end{figure}

The MD dataset provides training data for the derivatives of the entropy with respect to temperature and concentration. Knowing these derivatives, we reconstructed the absolute entropy by anchoring each phase at a reference point. To obtain this reference, we
carried out phonon calculations, evaluating the quantity
$\frac{1}{N} \log\det \hat{H}$,
which gives the entropy of the HCP and BCC phases at \(T = 0 K\).
We then employed Gaussian-process regression to extrapolate
\(\frac{1}{N}\log\det\hat{H}\) to the thermodynamic limit \(N\to\infty\),
as illustrated in Figure \ref{fig:hcp_ti_fit_log_det}.

We still require an anchor point for the liquid phase, which we obtain from the melting temperatures. 
At the melting point, the free energies of the solid and liquid phases are equal; because the solid free energies are already known, this equality furnishes the liquid anchor. 
To determine the melting temperatures, we employ the algorithm of Klimanova \textit{et al.}~\cite{Klimanova_2023}, which combines nonlinear Bayesian regression with finite-size extrapolation to the thermodynamic limit and provides a confidence interval for the resulting melting points. 
The workflow is illustrated in Figure \ref{fig:bcc_v_melting_point} for BCC titanium: first, melting temperatures and their uncertainties are estimated for several system sizes, and then Gaussian-process regression is used to extrapolate to \(N\!\to\!\infty\). 
We obtain
\(
T_{\mathrm{m}}(\mathrm{Ti}) = 1763 \pm 2 K\) and 
\( T_{\mathrm{m}}(\mathrm{V})  = 2180 \pm 2 K
\).
The vanadium value agrees well with the experimental melting point of \(2187~\text{K}\) \cite{80SMI}, whereas the titanium value is underestimated by roughly \(200~\text{K}\)—a common bias of DFT calculations with the PBE functional \cite{Rosenbrock2021, Andolina_2024} (experimental \(T_{\mathrm{m}}(\mathrm{Ti}) = 1941~\text{K}\) \cite{80SMI}).

\subsubsection{Free‐energy reconstruction}

\begin{figure*}[htbp]
    \centering
        \begin{subfigure}[t]{0.32\textwidth}
        \centering
        \includegraphics[width=\linewidth]{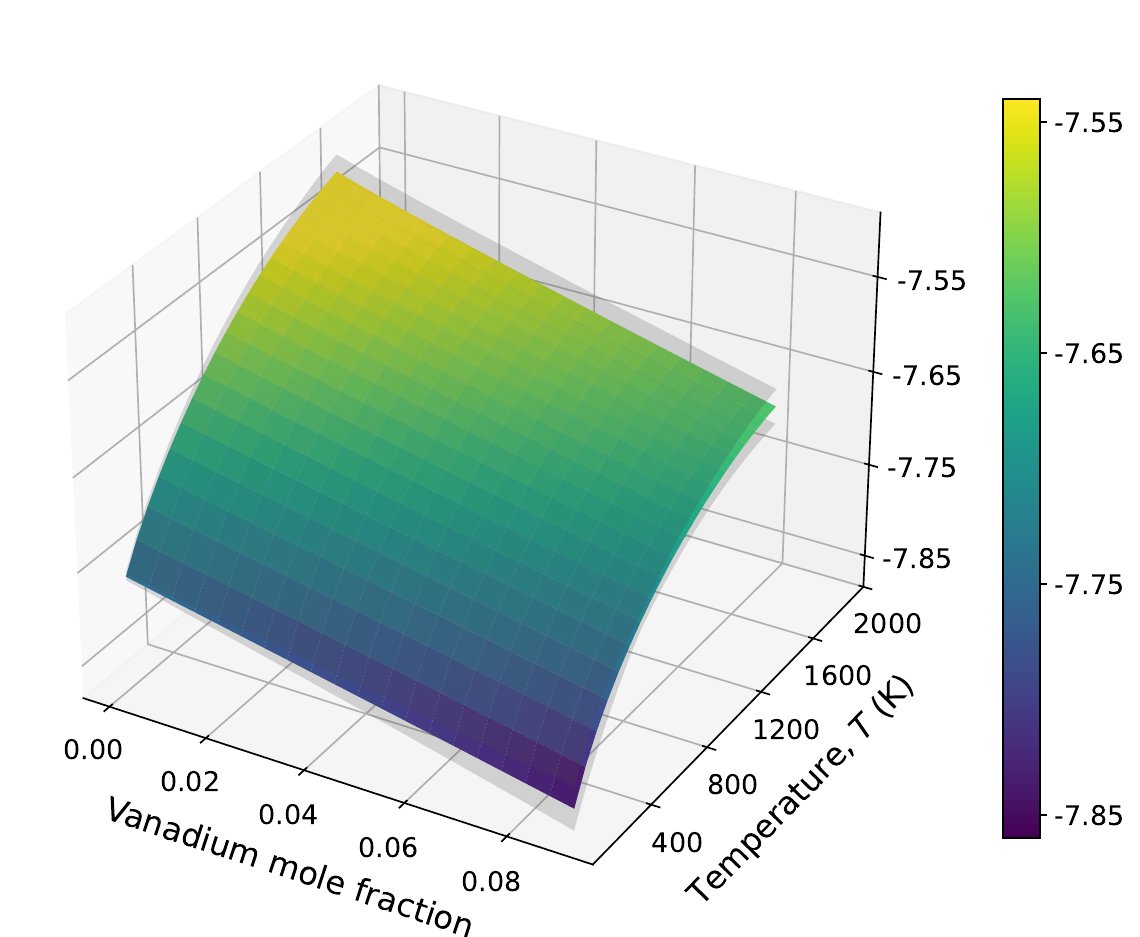}
        \caption{HCP phase}
    \end{subfigure}
    \hfill
    \begin{subfigure}[t]{0.32\textwidth}
        \centering
        \includegraphics[width=\linewidth]{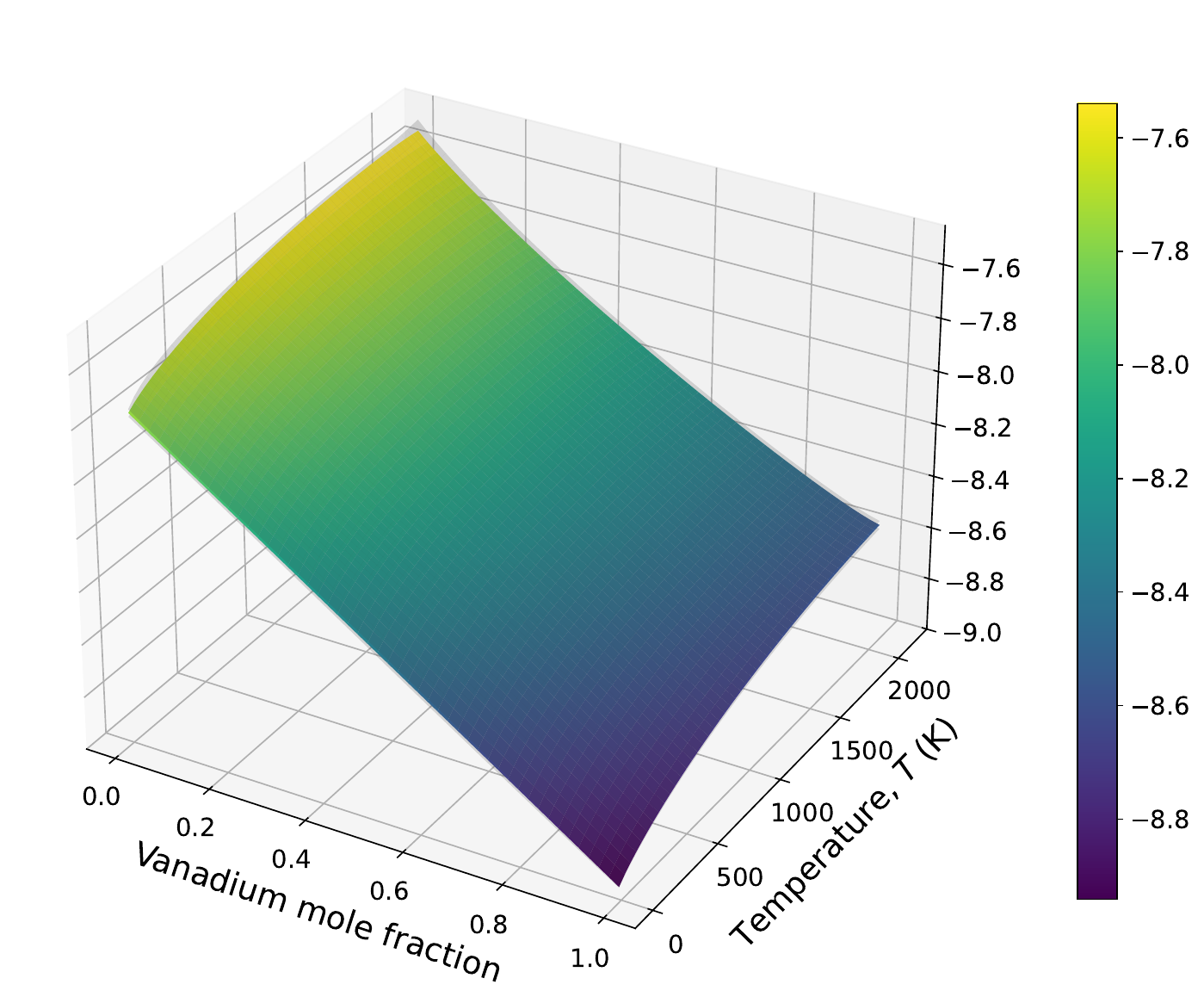}
        \caption{BCC phase}
    \end{subfigure}
    \hfill
    \begin{subfigure}[t]{0.32\textwidth}
        \centering
        \includegraphics[width=\linewidth]{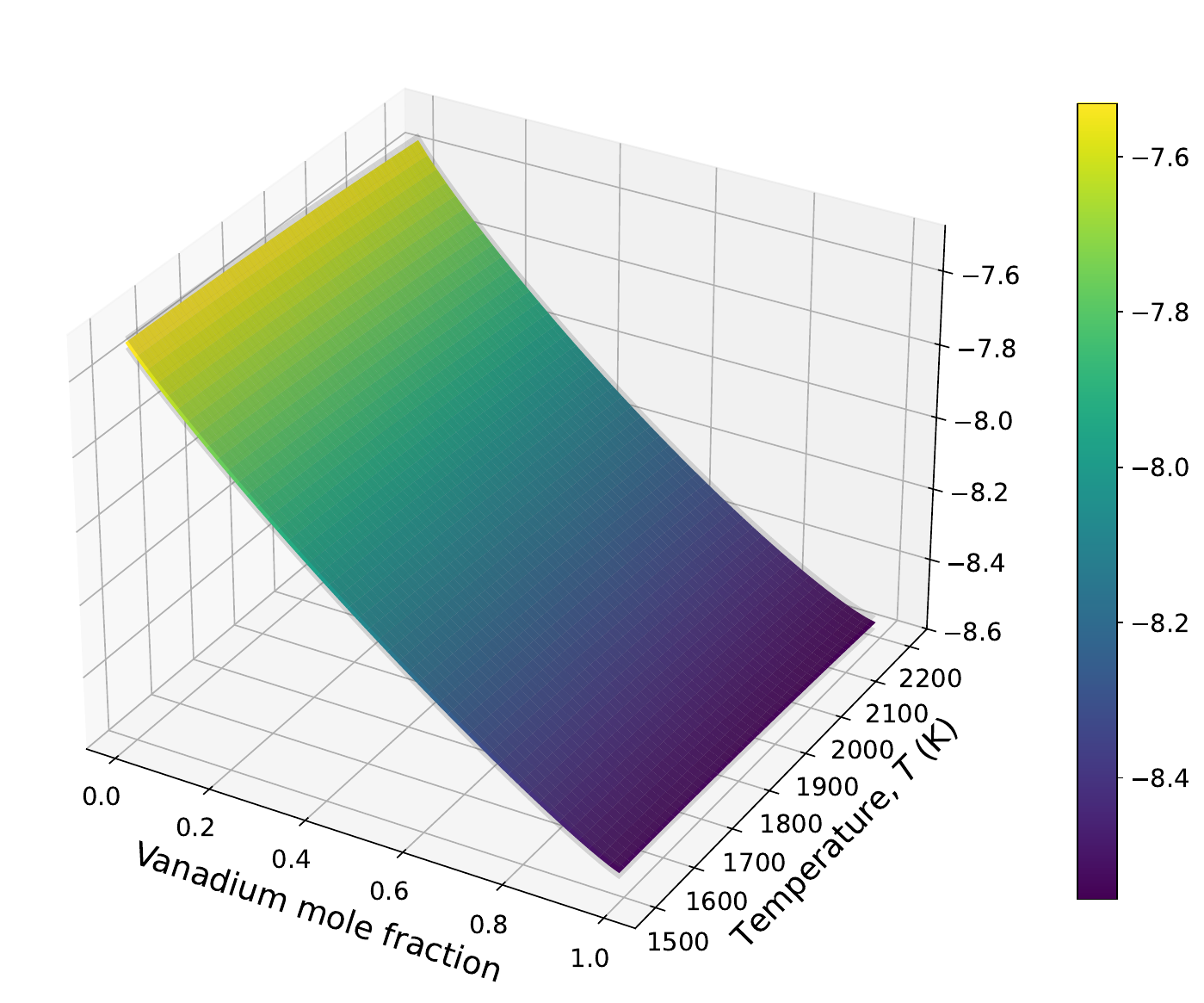}
        \caption{Liquid phase}
        \label{fig:sub_third}
    \end{subfigure}
    \caption{
    Free-energy surfaces of the HCP, BCC, and liquid phases, extrapolated to the thermodynamic limit ($N \rightarrow \infty$).
    The semitransparent grey envelopes indicate the associated 1-$\sigma$ confidence intervals.
    The reconstructed free-energy surfaces are subsequently used to infer the phase-transition boundaries.}
    \label{fig:free_energy_plots}
\end{figure*}

We trained the algorithm on a dataset that combined MD simulations, melting points from coexistence simulations, and phonon calculations for the HCP and BCC phases.  
The algorithm then reconstructed the free energies,
\(G = G(T,c)\), for each phase together with their confidence intervals, as shown in Figure \ref{fig:free_energy_plots}.

\begin{figure}[h]
\includegraphics[width=0.8\textwidth]{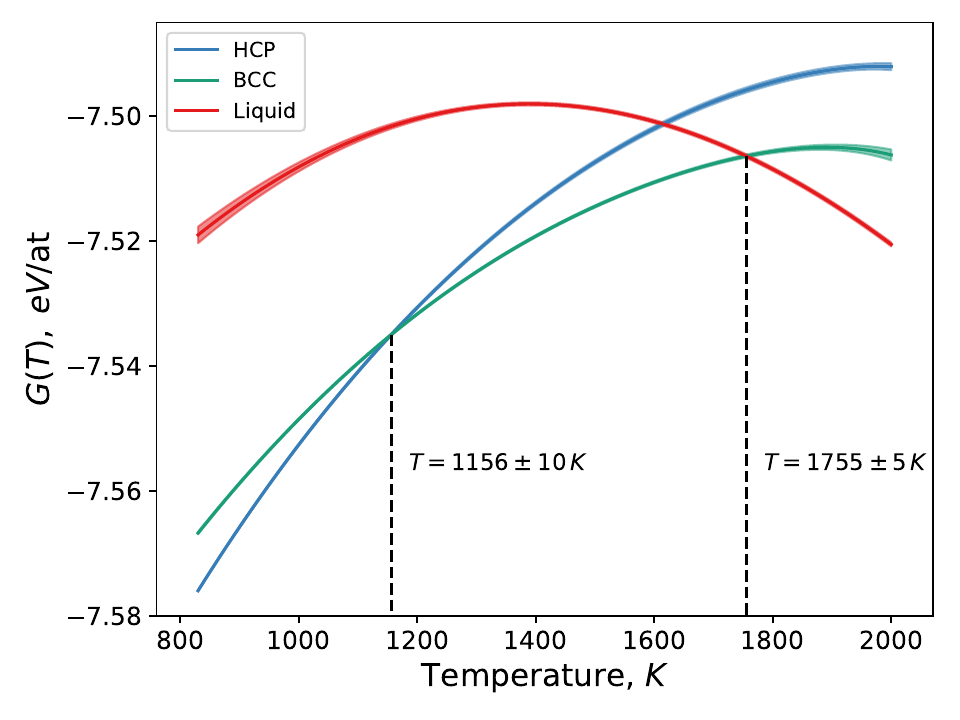}
\caption{Free-energy curves with 1-$\sigma$ confidence intervals for the BCC, HCP, and liquid phases of titanium, reconstructed from a reduced dataset in which the titanium melting point was withheld for validation.
Dashed lines mark the predicted phase-transition temperatures: the HCP\,$\rightarrow$\,BCC transition occurs at \(T_{\mathrm{HCP\!-\!BCC}} = 1156 \pm 10 K\), and melting at \(T_{\mathrm{melt}}(\mathrm{Ti}) = 1755 \pm 5 K\). 
The predicted melting temperature agrees, within two standard deviations, with the value obtained from previous coexistence simulations \(\bigl(T_{\mathrm{melt}}(\mathrm{Ti}) = 1762 \pm 2  K\bigr)\), thereby validating the algorithm.} 
\label{fig:free_energies_titanium}
\end{figure}

As an independent validation, we retrained the algorithm on a reduced dataset from which the titanium melting point was removed---retaining the vanadium melting point as the \emph{sole} anchor for the liquid phase. 
Using this dataset, the algorithm reconstructed the liquid free-energy surface and integrated it consistently from \(c = 1\) to \(c = 0\).
The resulting free-energy curve at \(c = 0\), together with those for the HCP and BCC phases, is shown in Figure \ref{fig:free_energies_titanium}.  
The intersections of these curves indicate titanium phase transitions: the HCP\,$\rightarrow$\,BCC transition occurs at \(T_{\mathrm{HCP\!-\!BCC}} = 1156 \pm 10 K\), and melting occurs at \(T_{\mathrm{melt}}(\mathrm{Ti}) = 1755 \pm 5 K\). 
The predicted titanium melting temperature agrees, within two standard deviations,
with the value obtained from previous coexistence simulations
\(\bigl(T_{\mathrm{melt}}(\mathrm{Ti}) = 1762 \pm 2 K\bigr)\),
demonstrating the accuracy and efficiency of the algorithm.

\subsubsection{Ti--V phase diagram}

\begin{figure*}[h!]
\includegraphics[width=1.0\textwidth]{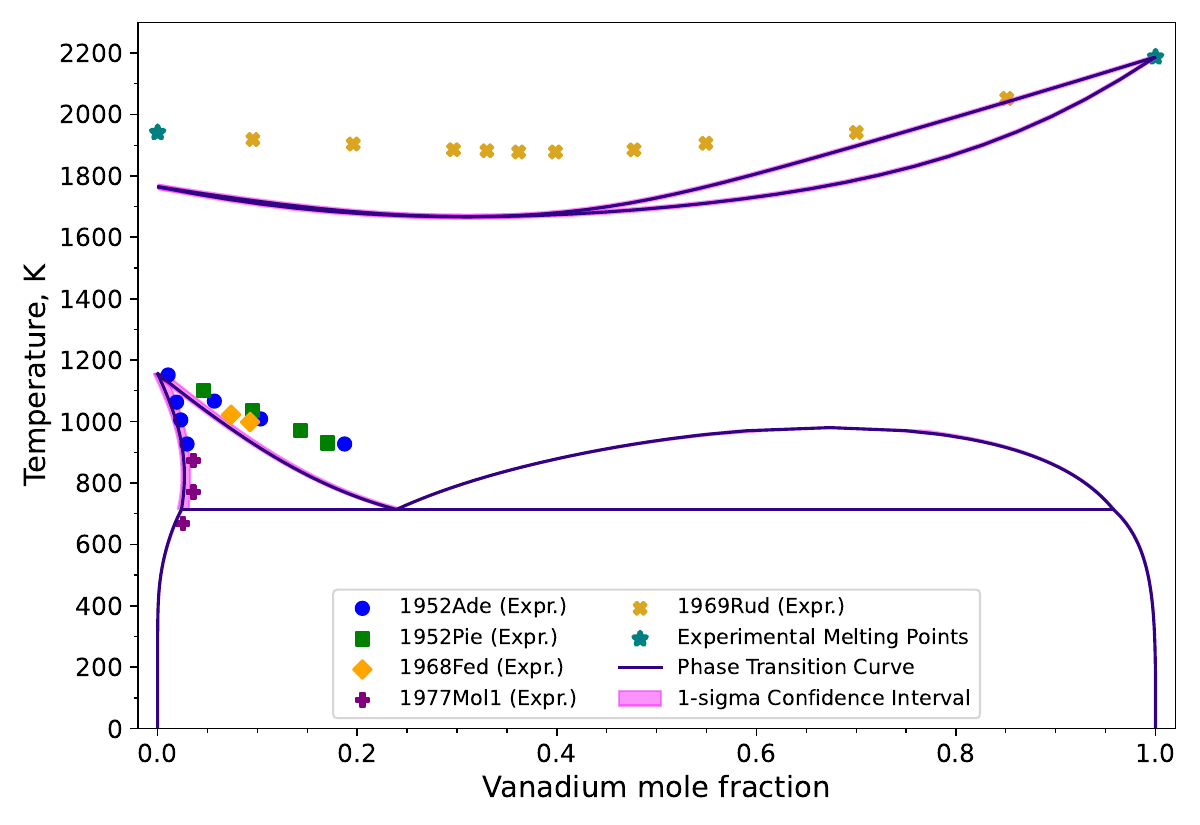}
\caption{Ti–V phase diagram with the $2\sigma$ confidence interval in the thermodynamic limit ($N \rightarrow \infty$) compared with experimental data from 1952Ade \cite{1952Ade}, 
1952Pie \cite{1952Pie}, 
1968Fed \cite{1968Fed}, 
1977Mol1 \cite{1977Mol1}, 
1969Rud \cite{1969Rud}. 
Experimental melting points for Ti and V are taken from 1969Rud \cite{1969Rud} and 80Smi \cite{80SMI}, respectively. 
The algorithm reproduces the Ti-rich boundary of the HCP–BCC two-phase region and qualitatively captures both the V-rich boundary and the solidus–liquidus envelope. 
The predicted HCP–BCC transition temperature,
\(T_{\mathrm{HCP\!-\!BCC}} = 1156 \pm 10 K\), matches the experimental value of \(1155 K\) within one standard deviation.
In the vanadium-rich composition range, where no experimental data are available, the algorithm predicts a BCC miscibility gap that terminates at a critical point of $T_\mathrm{crit} = 980 \pm 3 K$ and $c_\mathrm{crit} = 0.673 \pm 0.005$.
}
\label{fig:tiv_whole_wo_correction}
\end{figure*}

Our algorithm produced the phase diagram of the Ti–V binary system across the entire composition range, together with \(2\sigma\) confidence intervals in the thermodynamic limit \((N \rightarrow \infty)\), as shown in
Figure  \ref{fig:tiv_whole_wo_correction}. 
Experimental data available in the literature
\cite{1952Ade, 1952Pie, 1968Fed, 1977Mol1, 1969Rud, 80SMI} are plotted for comparison.  
(comparation with previous first principles computations
The algorithm reproduces the key experimental features: the Ti-rich boundary of the HCP–BCC coexistence region is matched quantitatively, whereas the V-rich boundary and the solidus–liquidus envelope are captured qualitatively. 
Moreover, the predicted HCP–BCC transition temperature,
\(T_{\mathrm{HCP\!-\!BCC}} = 1156 \pm 10 K\), matches the experimental value of \(1155 K\) within one standard deviation.

\begin{figure}[h!]
\centering
\includegraphics[width=0.8\textwidth]{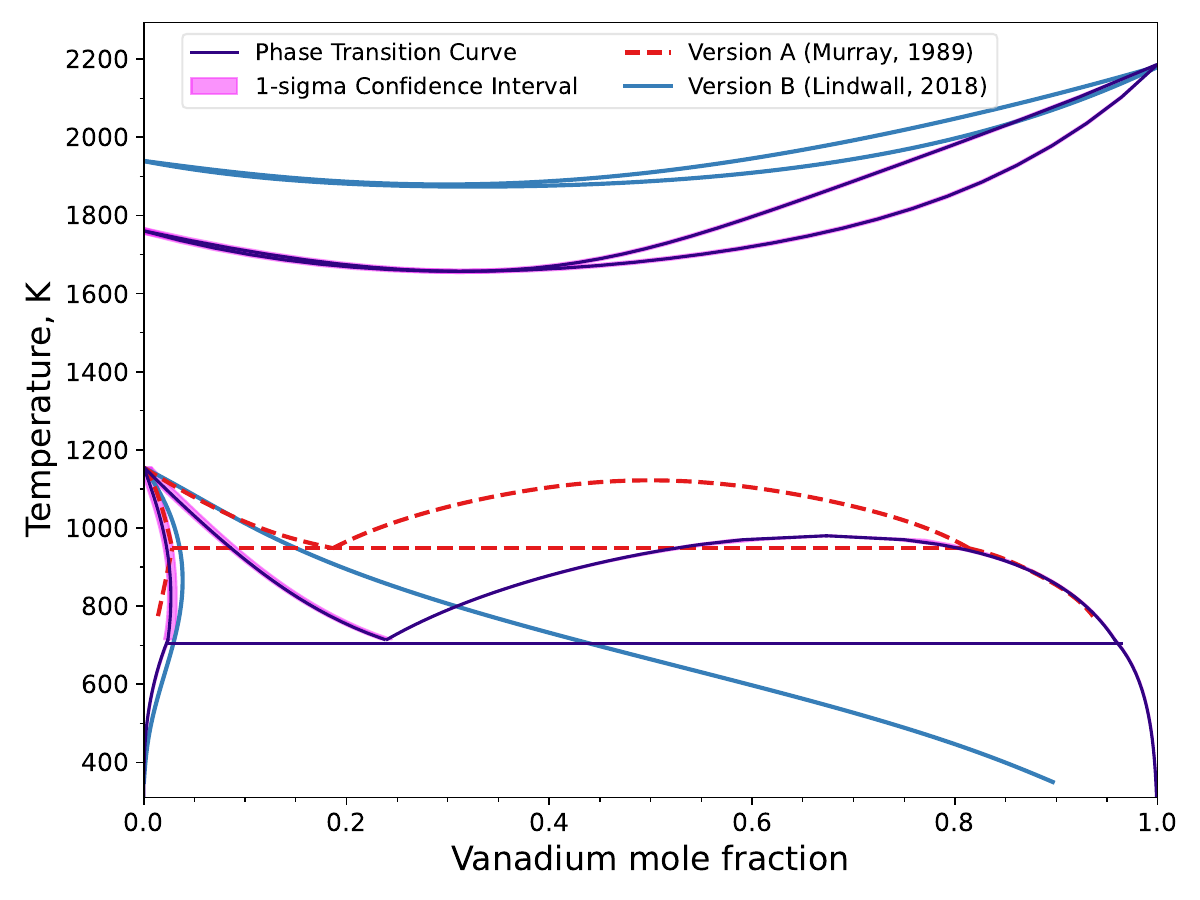}
\caption{
Two versions of the Ti–V phase diagram based on experimental data are reported in the literature: Version A features a miscibility gap, while Version B shows complete solid solubility. The phase diagram inferred by our Bayesian algorithm—based solely on quantum-mechanical calculations—predicts a miscibility gap, thereby supporting Version A.
}
\label{fig:compare_experiments_with_computational}
\end{figure}  

The Ti–V phase diagram predicted in this work exhibits a BCC miscibility gap in the V-rich region, for which no experimental data are currently available. 
The critical point is located at \(T_{\mathrm{crit}} = 980 \pm 3 K\) and \(c_{\mathrm{crit}} = 0.673 \pm 0.005\).
Notably, the phase diagram was generated entirely from quantum-mechanical calculations, with no experimental input for comparison with previous first-principles studies \cite{Sluiter1991, CHINNAPPAN2016125}, see \ref{sec:compare_computational_literature}).
This enables an unbiased comparison with the two experimental assessments reported in the literature (Figure \ref{fig:compare_experiments_with_computational}).
Our results favor the version A, which includes a BCC miscibility gap, over Version B, which do not.
Moreover, the present findings do not support the claim of Lindwall
\textit{et al.}~\cite{Lindwall2018} that oxygen impurities are responsible for the formation of the miscibility gap: the composition in our simulations was strictly controlled, and oxygen was not included.

\section{Summary and Concluding Remarks}

Two experimental assessments of the Ti–V phase diagram predominate in the literature: Version A reports a miscibility gap  (e.g., Murray 1989 \cite{murray1989phase}), whereas Version B shows no such gap (e.g., Lindwall 2018 \cite{Lindwall2018}); both are shown in Figure 9.
Here, we constructed the Ti--V phase diagram over the full temperature–composition range entirely from quantum-mechanical calculations.  
A molecular-dynamics dataset for the BCC, HCP, and liquid phases—supplemented by zero-kelvin phonon calculations and melting points of Ti and V obtained from coexistence simulations—served as input to our Bayesian learning framework.  
The algorithm propagates the uncertainty in the input data to the phase-transition boundaries and extrapolates them to the thermodynamic limit ($N \!\to\! \infty$), yielding an uncertainty-quantified Ti–V phase diagram.   
When compared with the available experimental data, our algorithm quantitatively reproduced the Ti-rich boundary of the HCP–BCC two-phase region and qualitatively captures both the V-rich boundary and the solidus–liquidus envelope. 
The predicted HCP–BCC transition temperature, \(T_{\mathrm{HCP\!-\!BCC}} = 1156 \pm 10 K\), matches the experimental value of \(1155 K\) within one standard deviation.

The resulting phase diagram supports Version A, which features a miscibility gap. Relative to Murray (1989), our prediction places the gap about $250 K$ lower in temperature; the critical point is likewise shifted — by about $250 K$ in temperature and $0.15$ in composition — to $T_{\mathrm{crit}} = 980 \pm 3\ \mathrm{K}$ and $c_{\mathrm{crit}} = 0.673 \pm 0.005$.
We find no evidence that oxygen impurities are required for miscibility gap formation, contradicting the argument put forward by Lindwall \textit{et al.}\ \cite{Lindwall2018}.

Beyond resolving this specific discrepancy, our workflow highlights particular compositions and temperatures where experiments would be most informative. 
For example, our predictions place Ti-70V at 800 K within the miscibility gap region, indicating that targeted experiments on high-purity samples at this composition would provide a direct test of the contested feature of the Ti–V system.

Looking ahead, we aim to extend our ab initio, uncertainty-aware Bayesian framework to ternary alloys, enabling, in particular, direct quantification of oxygen’s influence on Ti–V phase stability and explicit treatment of oxygen-bearing phases.


\section*{Acknowledgments}
This work was supported by Russian Science Foundation (grant number 23-13-00332, https://rscf.ru/project/23-13-00332/). 

\section*{Data Availability}
The data that support the findings of this study are openly available in ``Ti–V Moment Tensor Potential Dataset'' at https://doi.org/10.5281/zenodo.15755826, reference number \cite{miryashkin_2025_15755826}. The repository provides both the reference dataset and the fitted MTP for the titanium–vanadium binary alloy. All reference energies and forces were generated with density functional theory calculations, while the potential itself was trained via active learning.

\appendix

\section{Comparison of First-Principles Ti-V Phase Diagrams}
\label{sec:compare_computational_literature}

\begin{figure}[H]
\centering
\includegraphics[width=0.8\textwidth]{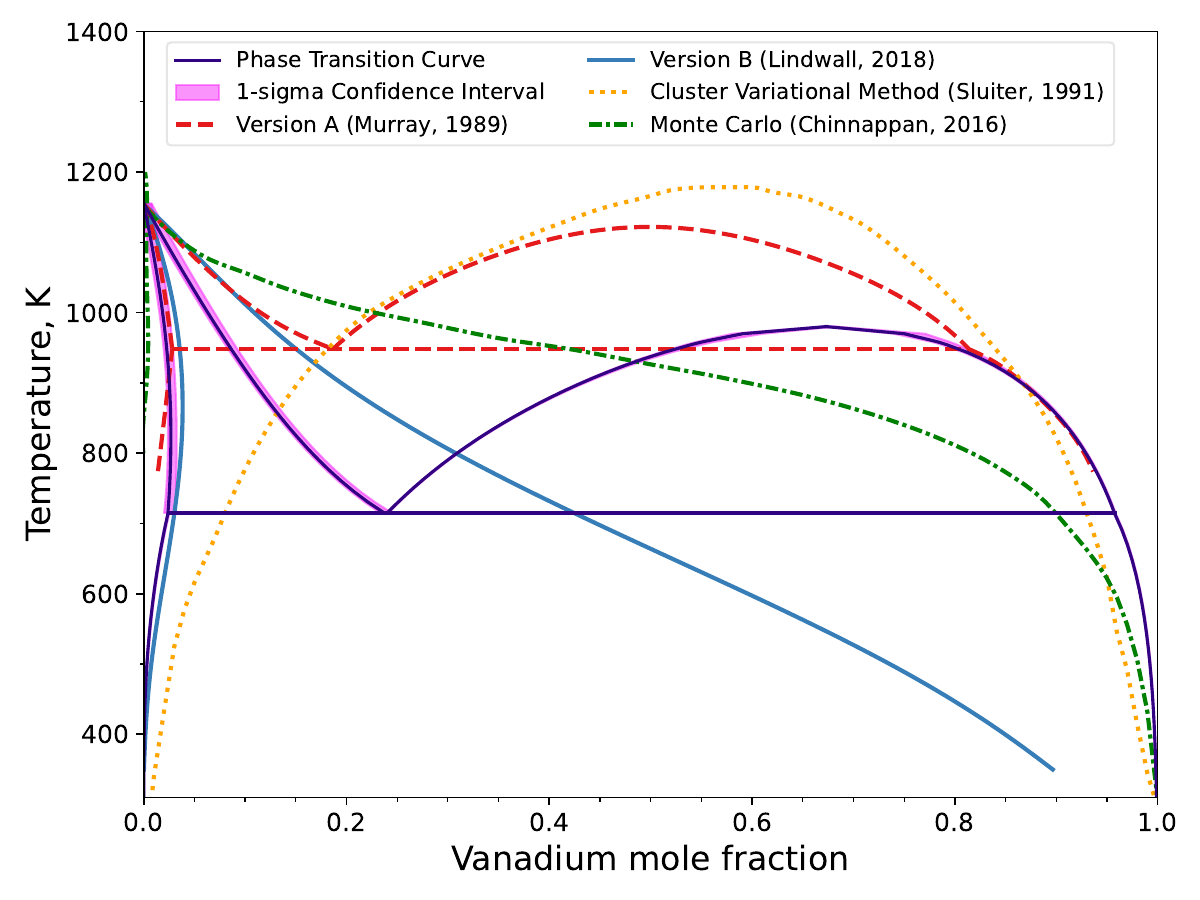}
\caption{Computed phase diagram of Ti-V compared to two interpretations of experimental data (Version A and Version B). The plot also presents two theoretical studies, both utilizing ab initio computations to parametrize the interactions in the system. Sluiter \cite{Sluiter1991} utilizes the cluster variation method to compute the free energy of the solid phase, while Chinnappan \cite{CHINNAPPAN2016125} uses Monte Carlo for compositional sampling (along with the bond stiffness versus bond length method to capture the vibrational contribution).}
\label{fig:compare_computational_literature}
\end{figure} 

\bibliographystyle{elsarticle-num} 
\bibliography{refs} 





\end{document}